\documentclass[]{aa}      

\usepackage{graphicx}
\usepackage{linenoaa}
\linenumbers

\usepackage{subcaption}
\usepackage{soul}
\usepackage{txfonts}
\usepackage{multirow}
\usepackage{natbib}
\usepackage{placeins}
\usepackage{bm} 
\usepackage{upgreek}
\usepackage{overpic}
\usepackage{natbib}
\usepackage{caption}
\usepackage{comment}
\usepackage{rotating}
\usepackage{pdflscape}
\usepackage{subcaption} 
\usepackage{placeins}
\usepackage{flafter} 
\usepackage{float}
\usepackage[maxfloats=256]{morefloats}
\usepackage[bookmarks=false, colorlinks=true, citecolor=blue, linkcolor=blue]{hyperref}
\usepackage[normalem]{ulem}
\def\kms{\,km\,s$^{-1}$\,}

\begin{document} 

\title{Discovery of synchronized periodic variability of methanol maser features in G26.598$-$0.024
}

 \author{P. Wolak
          \inst{1} \href{https://orcid.org/0000-0002-5413-2573}     {\includegraphics[scale=0.5]{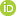}}
          \and
          M. Szymczak
          \inst{1} \href{https://orcid.org/0000-0002-1482-8189}
          {\includegraphics[scale=0.5]{orcid.png}}
          \and    
          A. Kobak
          \inst{1} \href{https://orcid.org/0000-0002-1206-9887}{\includegraphics[scale=0.5]{orcid.png}}
          \and
          A. Varma
          \inst{1} \href{https://orcid.org/0009-0000-0076-1123}
          {\includegraphics[scale=0.5]{orcid.png}}
          \and
          A. Bartkiewicz
          \inst{1} \href{https://orcid.org/0000-0002-6466-117X}{\includegraphics[scale=0.5]{orcid.png}}
          \and
          J. O. Chibueze 
          \inst{2,3}
          \href{https://orcid.org/0000-0002-9875-7436}
          {\includegraphics[scale=0.5]{orcid.png}}
          \and
          B. Stecklum
          \inst{4} 
          \and
          C. J. Ugwu
          \inst{2} 
          \href{https://orcid.org/0000-0001-8467-6582}
          {\includegraphics[scale=0.5]{orcid.png}}
          } 
  
   \institute{Institute of Astronomy, Faculty of Physics, Astronomy and Informatics, Nicolaus Copernicus University, Grudziadzka 5, 87-100 Torun, Poland
   \and
   Department of Mathematical Sciences, University of South Africa, Cnr Christian de Wet Rd and Pioneer Avenue, Florida Park, 1709 Roodepoort, South Africa
   \and
   Department of Physics and Astronomy, Faculty of Physical Sciences, University of Nigeria, Carver Building, 1 University Road, Nsukka 410001, Nigeria
    \and
    Thüringer Landessternwarte Tautenburg, Sternwarte 5, 07778 Tautenburg, Germany
}

\date{Received 16 June 2025 / Accepted 18 July 2025 }

  \abstract
{}
{We report the discovery of periodic maser variability of an unusual pattern in the high-mass young stellar object G26.598$-$0.024.}
{A ten-year monitoring of the 6.7\,GHz methanol maser was carried out with the Torun 32\,m radio telescope. The archival data collected so far were also used to characterize the target with high angular resolution and examine its infrared variability.}
{We found anticorrelated flux variations of the opposite blueshifted and redshifted emission features with a period of 70.1$\pm$2.2\,d and the relative amplitude of 1.3 and 0.6, respectively. The light curves are best fit with a sinusoidal function modulated by mild changes in the average flux density on 3-5\,yr timescales. The emission of the middle parts of the spectrum shows only long-term variability. High-angular-resolution data indicate that the maser is associated with one of the two 1.3\,mm dust emission cores, while the periodic emission comes from two extended regions $\sim$1500\,au apart. We discuss several possible causes of the peculiar variability.}
{}


\titlerunning{Synchronized periodic variability of maser features}
\authorrunning{P. Wolak et al.}

\maketitle

\section{Introduction}
Long-term observations of the 6.7\,GHz methanol line reveal growing evidence for periodic variation of the maser emission toward high-mass star-forming regions (e.g. \citealt{goedhart2004,goedhart2014}; \citealt{fujisawa2014};\citealt{maswanganye2015,maswanganye2016}; \citealt{szymczak2015,szymczak2018a}; \citealt{sugiyama2017}; \citealt{olech2019,olech2022}; \citealt{tanabe2023}). Periodic changes in the maser flux of all or some of the spectral features can be explained by periodic increases in the accretion rate clocked in the binary system, which causes an increase in luminosity and, consequently, an increase in the flux of photons pumping in the infrared range (\citealt{araya2010}; \citealt{parfenov2014}; \citealt{munoz2016}) or, due to the influence of stellar winds, an increase in the level of radio radiation backgrounds (\citealt{van_der_walt2011,van_der_walt2016}). It is also postulated that massive stars with high accretion rates may periodically change their luminosity due to the $\kappa$-mechanism (\citealt{inayoshi2013}). In the maser sources reported so far, changes in the flux of spectral features are synchronized with each other or, at most, show time delays resulting from differences in the geometric paths of the pumping photons (\citealt{goedhart2014}; \citealt{maswanganye2016}; \citealt{olech2019}).

Here, we present 6.7\,GHz methanol maser observations toward the massive star-forming region G26.598$-$0.024 (hereafter G26), also known as IRAS18372$-$0541. The near kinematic distance of the target is 1.7\,kpc (\citealt{reid2019}). A bright (sub)millimeter condensation with an extended elliptical-shaped halo $\gtrsim$15\arcsec and $\gtrsim$45\arcsec was detected in ATLASGAL (\citealt{schuller2009}) and BOLOCAM (\citealt{rosolowsky2010}) surveys, respectively. 
{\it Spitzer} data revealed a core-halo nebula ($\sim$5\arcsec) at 5.8 and 8.0\,$\mu$m (\citealt{benjamin2003}). Gemini observations at 8 and 18\,$\mu$m revealed a double structure of a central object with peaks separated by 0\farcs97 along a position angle of 45$\degr$ (\citealt{bartkiewicz2010}). The 6.7\,GHz methanol and 22.2\,GHz water masers are associated with the NE component. In contrast, the SW component coincides with the H\,\textsc{ii} region (\citealt{bartkiewicz2009, hu2016}) of a spectral index of 0.23 between 1.4 and 5\,GHz (\citealt{becker1994}). 6.7\,GHz maser monitoring at a one-month cadence showed weak variations on timescales of less than four years, except for the feature near 23.4\kms (\citealt{szymczak2018a}). However, inspection of single-dish data from the literature suggests significant variability in maser emission over three decades (e.g., \citealt{schutte1993,codella2000,szymczak2000,fontani2010,breen2015,yang2019}). High-cadence observations since March 2023, which are reported in the paper, clearly show evidence of anticorrelated periodic variability in the blueshifted and redshifted emission.

\section{Observations}
\subsection{32\,m telescope}
The methanol maser line at 6.7\,GHz associated with G26 was observed from June 2009 to February 2013 as part of the monitoring project of a large sample of HMYSO performed with the Torun 32-m radio telescope (\citealt{szymczak2018a}). Further observations of this relatively faint, low-variability source were suspended for more than three years because of telescope time constraints. The observations resumed in December 2017 and continued until March 2023, with a median cadence of 25 days, and then, on average, every other day until the present. The spectral resolution was 0.09\kms, and a typical 1$\sigma$ rms noise level was 0.20 to 0.25\,Jy. The telescope parameters and observing strategy were presented in \cite{szymczak2018a}. 

\subsection{Ancillary data}
G26 was observed in the 6.7\,GHz methanol line using the European VLBI network (EVN) in June 2007 (MJD 54264, \citealt{bartkiewicz2009}). The synthesized beam was $6\times12$\,mas$^2$, the velocity resolution was 0.18\kms, and the $1\sigma_{\mathrm{rms}}$ noise per channel was 4\,mJy\,beam$^{-1}$. We used the parameters of the measured spots available in \cite{bartkiewicz2009} to obtain the overall structures of maser cloudlets, where a cloudlet is defined as a group of maser spots appearing in three contiguous spectral channels and coinciding in position within half of the synthesized beam. 

A similar procedure was applied to the VLA data obtained in the C configuration in March 2012 (MJD 56004, \citealt{hu2016}). The synthesized beam, the spectral resolution, and the $1\sigma_{\mathrm{rms}}$ noise were $3\arcsec\times6\arcsec$, 0.18\kms and 33\,mJy\,beam$^{-1}$ per channel, respectively. The target was observed in August 2014 (MJD 56882) with the VLA (D configuration) as part of the GLOSTAR survey (\citealt{nguyen2022}). A circular synthesized beam of 18\arcsec\, in diameter was used and $\sigma_{\mathrm{rms}}$ noise of $\sim$18\,mJy\,beam$^{-1}$ was reached for 0.18\kms spectral resolution. We used the spots reported in \cite{nguyen2022}; the cross-correlated spectrum implies that they represent the peaks of cloudlets.

We retrieved ALMA 12\,m archival observations in Band 6 (project 2021.1.00311.S). Data were taken in August 2022 (MJD 59805) in three spectral windows: $\sim$216.6$-$218.5\,GHz, $\sim$219.9$-$220.8\,GHz and $\sim$230.0$-$233.7\,GHz with a uniform resolution of 0.977\,MHz (1.3\kms). The continuum emission was imaged using a Briggs weighting (robust 0.5) to optimize the angular resolution and sensitivity, averaging over all spectral windows after including only line-free regions with a resulting synthesized beam of $0\farcs30\times0\farcs28$ (PA = 67\degr) and the $\sigma_{\mathrm{rms}}$ noise of 0.3\,mJy\,beam$^{-1}$. The calibration and imaging of the data were performed using the common astronomy software application (CASA) version 5.4.0 (\citealt{casa2022}). The typical rms noise level of the spectral line data was 3$-$4\,mJy\,beam$^{-1}$ per 1.3\kms channel. In this report, we focus only on the transitions of CO (2$-$1, 230.538\,GHz), SiO (5$-$4, 217.105\,GHz) and CH$_3$OH ($4_{(2,2)} - 3_{(1,2)}$ E, 218.440\,GHz) as typical outflow and disk tracers, respectively.
    
The image of the target at 8.6$\,\mu$m was taken with Gemini South with $\sim0\farcs15$ resolution using the technique and procedures described in \cite{debuizer2012}. We acquired the map as published in \cite{bartkiewicz2010}.

W1 (3.4\,$\mu$m) and W2 (4.6\,$\mu$m) photometry were downloaded from the NEOWISE (\citealt{mainzer2014}) single-exposure database. The high-quality (ph$_{-}$qual = AA) exposures were only used to obtain average magnitudes at 21 epochs since April 2014 (MJD 56750).

\section{Results and analysis}\label{sec:results}
\begin{figure}
\centering
\begin{tabular}{cc}
\begin{overpic}[width=0.48\textwidth]{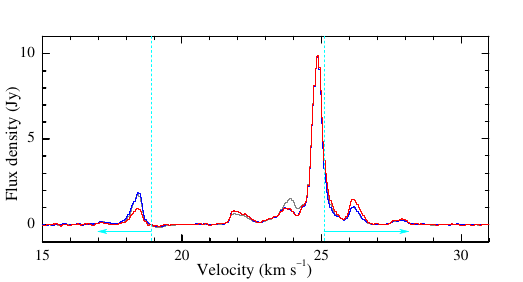}\put(12,39){(a)}\end{overpic} \\
        \begin{overpic}[width=0.46\textwidth]{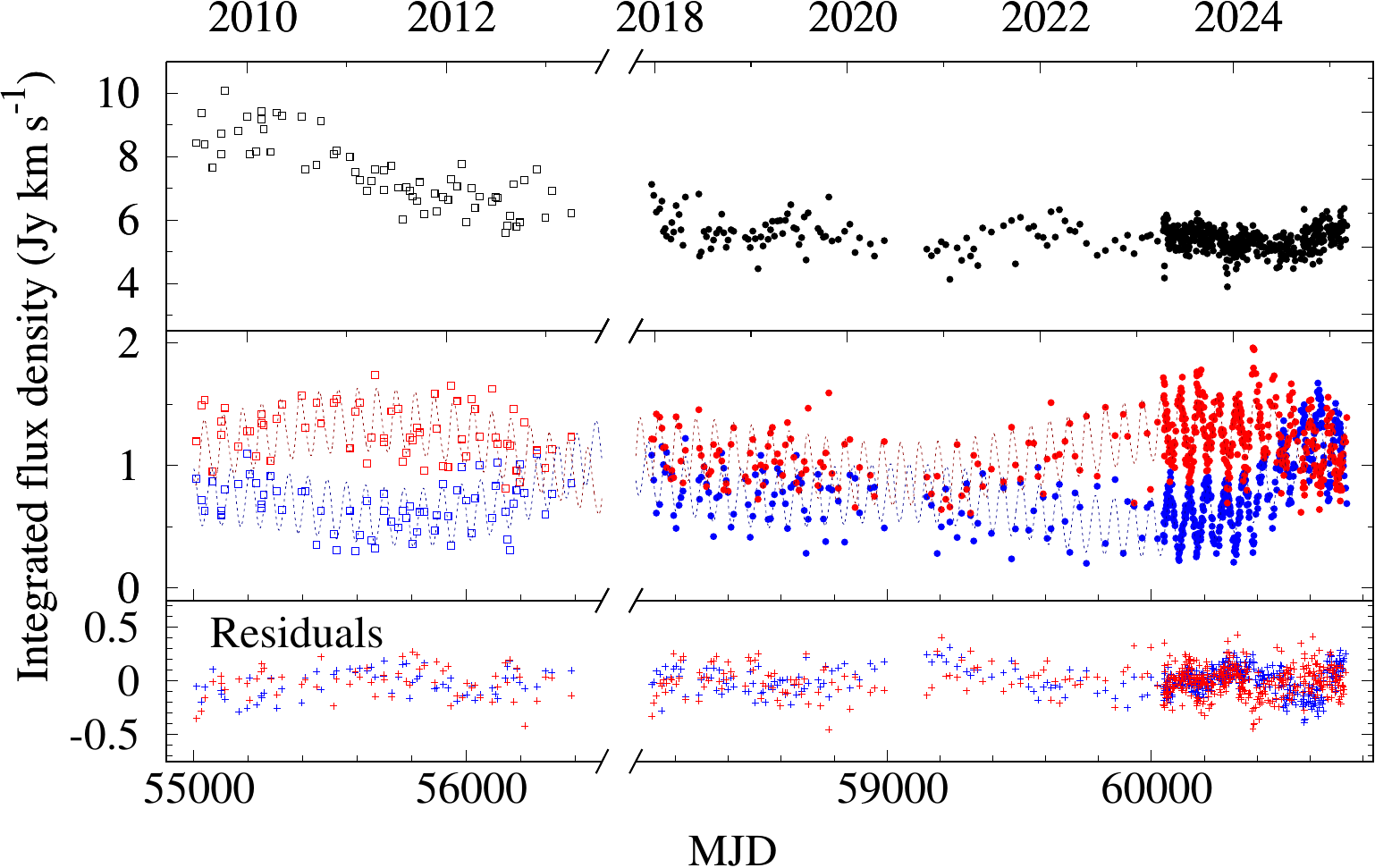}\put(10, 60){(b)}\end{overpic} \\
\end{tabular}
\caption{({\bf a}) Spectra of G26 averaged over three-week intervals around five consecutive maxima (blue) and minima (red) of the 18.4\kms\, feature for MJD 60045-60407 observations. The gray line shows the spectrum averaged over 480 observations from MJD 55010 to 60580. The cyan lines and arrows denote the velocity ranges in which the emissions exhibit anticorrelated variability. 
({\bf b}) Time series for velocity-integrated flux density in three velocity ranges: 18.0$-$18.7\kms (blue), 21.7$-$25.1\kms (gray), and 25.1$-$26.6\kms (red). Filled symbols represent the new data, and the data from \cite{szymczak2018a} are open. A sinusoidal function fits the blueshifted and redshifted emission light curves. 
   \label{fig:g26-spect-mm-all}}
\end{figure}

\begin{figure}
\centering
\begin{tabular}{cc}        
        \begin{overpic}[width=0.47\textwidth]{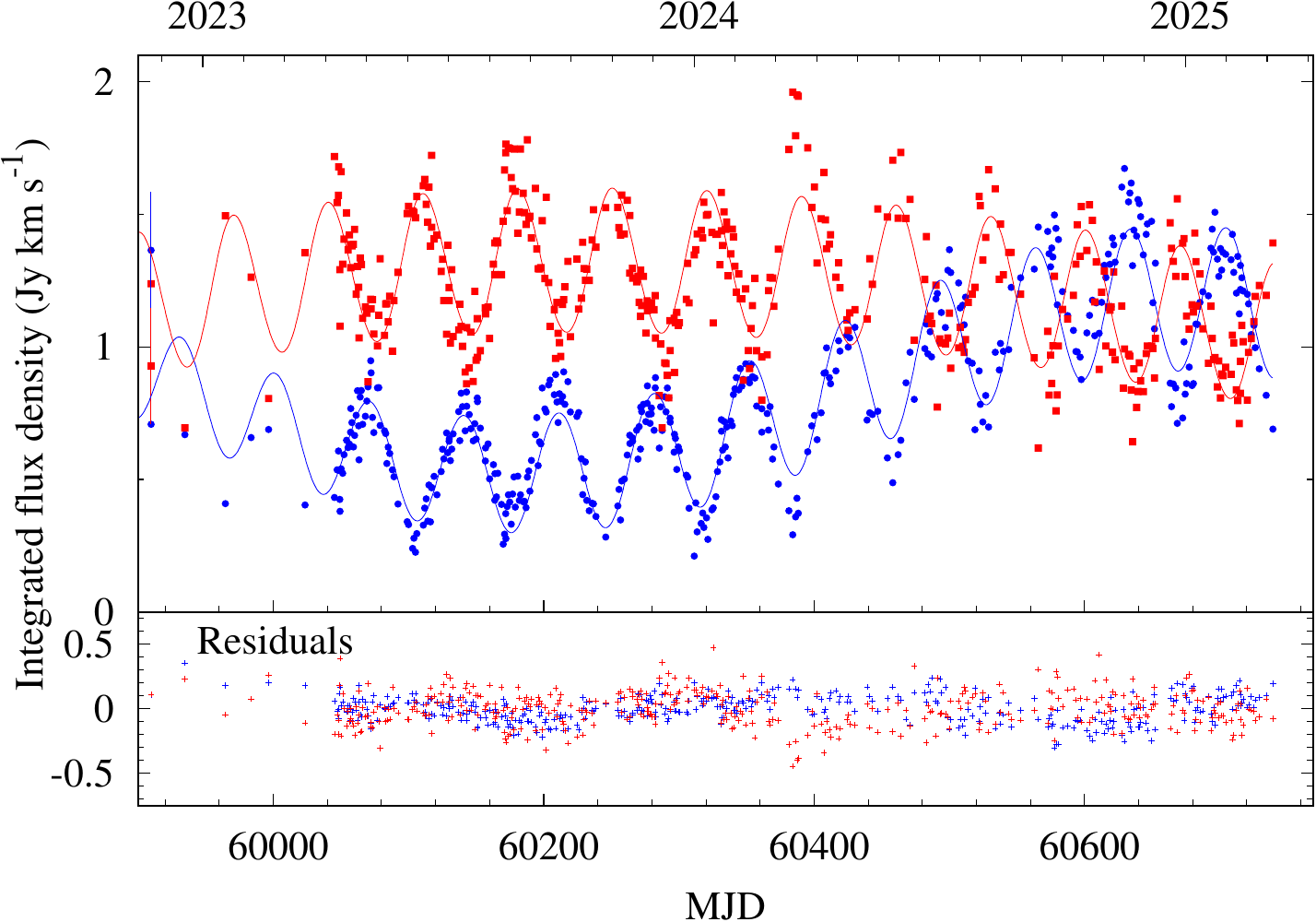}\put(14, 58){(a)}\end{overpic} \\
        \begin{overpic}[width=0.5\textwidth]{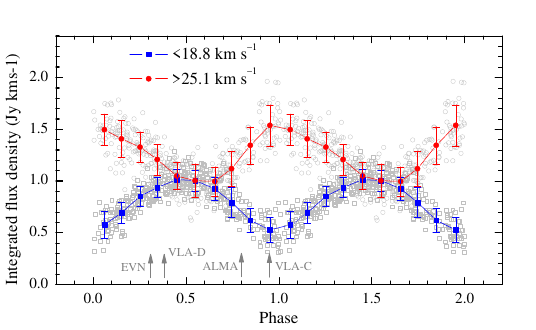}\put(15, 48){(b)}\end{overpic} \\
\end{tabular}
\caption{({\bf a}) Zoomed-in view of the light curves from Fig.\,\ref{fig:g26-spect-mm-all} during the high-cadence observations (MJD 60045-60739).
({\bf b}) Time series of the integrated emission at a velocity lower than 18.8\kms (squares) and higher than 25.1\kms\, (circles) folded modulo 70.1\,d. The filled symbols represent data binned with a 0.1 phase increment. The bars denote the standard deviation. The gray arrows mark the phase of EVN (\citealt{bartkiewicz2009}), VLA-C (\citealt{hu2016}), VLA-D (\citealt{nguyen2022}) and ALMA observations.}
\label{fig:g26-part-lc}
\end{figure}

The spectra of the target's 6.7\,GHz maser line in different periods and light curves are shown in Fig.\,\ref{fig:g26-spect-mm-all}. Two prominent blueshifted and redshifted spectral features, peaked at 18.4\kms and 26.2\kms, respectively, exhibit distinctly different amplitudes in the selected intervals around the maximum and minimum of the former feature. The emission peaked near 23.9\kms, showing a decrease in intensity, mainly due to the decrease by a factor of about four in the first four years of the observations \citep{szymczak2018a}. A closer look at the intensity changes in the individual spectral channels revealed that the emission exhibits periodic changes at velocities lower than 18.8 and greater than 25.1\kms. The light curves of the blueshifted ($<$18.8\,km\,s$^{-1}$; hereafter BS) and redshifted ($>$25.1\,km\,s$^{-1}$; hereafter RS) emission after MJD 59909 are enlarged in Fig.\,\ref{fig:g26-part-lc}. The periodic anticorrelated variations of the velocity-integrated flux density in both spectral parts are visible, while a sine function best fits the time series. The Lomb-Scargle (LS) periodogram \citep{astropy:2018} implies the period ($P$) of 70.1$\pm$2.2\,d. Here, the uncertainty is equal to half the width at half the maximum of the Gaussian fitting to the LS maximum. BS and RS emission show some anti-cyclic beating on a timescale of $\gtrsim$4\,yr. 

The phased light curves of the BS and RS emission are shown in Fig.\,\ref{fig:g26-part-lc}b. A low-order polynomial fitting removed $\gtrsim$4\,yr variations before folding the curve. The data phase-folded with the 70.1\,d period show that the relative amplitude of the variations is 1.3 and 0.6 for the BS and RS emission, respectively. In other words, G26 varies with the amplitude of 36\% and 23\%, the mean integrated flux density of the BS and RS parts of the spectrum, respectively. The ratio of rise-to-decay time of the flare is 0.57 and 0.35 for the BS and RS light curves, respectively. The BS emission peaked 33\,d (0.47$P$) earlier than the RS emission, while the time delay in the corresponding minima is 48\,d (0.68$P$). The two redshifted features, 25.2 and 26.2 km/s, were strong enough for us to attempt to measure the time delays. The time delay of these features is 1.7$\pm$0.8\,d (Fig.\,\ref{fig:g26-timelag}), which corresponds to the difference in light travel time over 290\,au.

\begin{figure*}[h]
\begin{minipage}[t]{0.48\textwidth}
  \begin{overpic}[width=\linewidth]{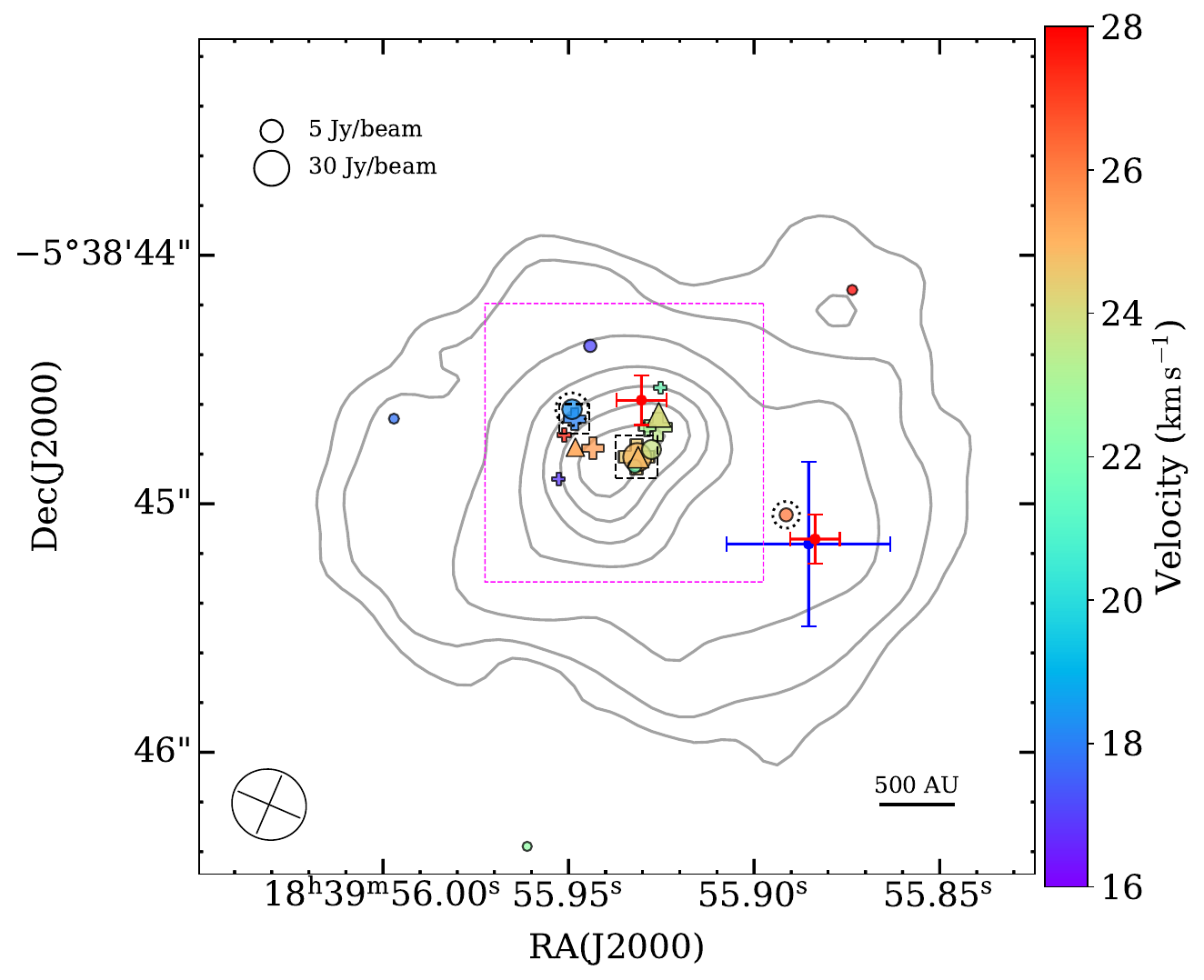}
   \put(19,74){(a)} 
  \end{overpic}
\end{minipage}%
\hfill
\begin{minipage}[t]{0.5\textwidth}
  \begin{overpic}[width=\linewidth]{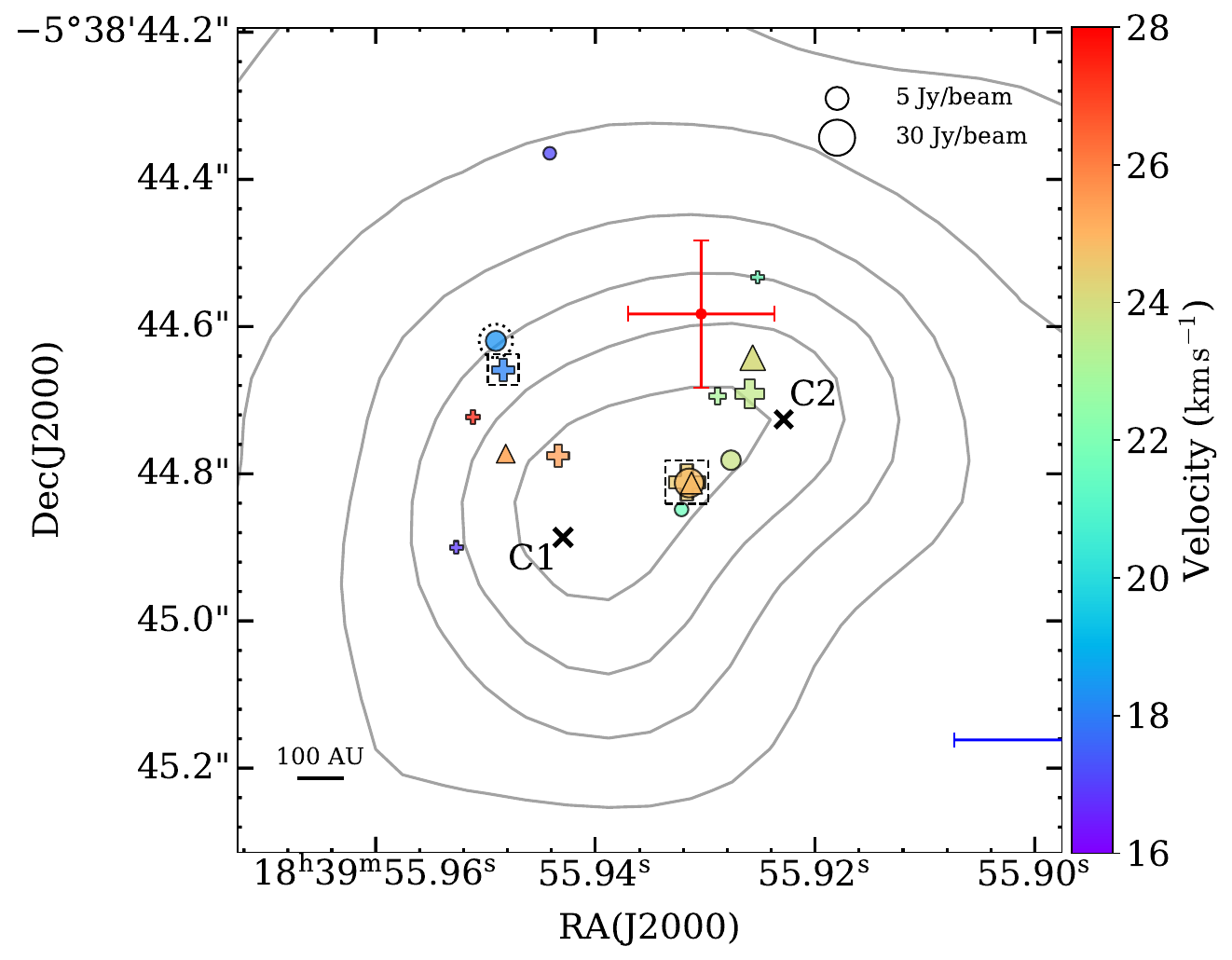}
    \put(22,71){(b)}
  \end{overpic}
\end{minipage}
\caption{({\bf a}) Spatial distribution of centroids of 6.7\,GHz methanol maser cloudlets overlaid with the 1.3\,mm ALMA continuum. EVN (\citealt{bartkiewicz2009}), VLA-D (\citealt{nguyen2022}), and VLA-C (\citealt{hu2016}) data are denoted with triangles, circles and crosses, respectively. The two crosses highlighted by dashed squares and two circle symbols surrounded by dotted circles indicate the strongest well-separated components in the spectrum, which exhibit periodic anticorrelated emission. The size of the symbol is proportional to the brightness of the cloudlet on a logarithmic scale, and its color represents the velocity shown by the wedges. The red and blue dots mark the positions with error bars of the 8.6\,$\mu$m (\citealt{bartkiewicz2010}) and 8.4\,GHz (\citealt{bartkiewicz2009}) sources, respectively. Contour levels are 3, 5, 10, 20, 30, 40, 50 times 1$\sigma_{\mathrm{rms}}$ of the 0.3\,mJy\,beam$^{-1}$. The synthesized beam is shown in the bottom left. ({\bf b}) Magnified view of central region marked by the dashed magenta box in (a). Crosses labeled C1 and C2 represent the 1.3 mm continuum cores listed in Table\,\ref{tab:table1}.
   \label{fig:g26-spots}}
\end{figure*}

The NEOWISE measurements at 3.4 and 4.6\,$\mu$m superimposed on the velocity-integrated maser flux density are summarized in Fig.\,\ref{fig:g26-neowise}. There was a general trend of brightening by $\sim$0.15 and $\sim$0.30\,magnitudes in 3.4 and 4.6\,$\mu$m bands. These NEOWISE light curves are a superposition of emission from at least two objects separated by 0\farcs97 (Fig.\,\ref{fig:g26-spots}). After removing long-term variations and using the ephemeris of the maser BS emission maxima, no variability with an amplitude greater than 0.035 magnitudes in band W2 was found.

EVN data taken at phase 0.31 (Fig.\,\ref{fig:g26-part-lc}b) revealed that the maser emission in G26 comes from an $\sim0\farcs36\times0\farcs15$ ($\sim630\times270$\,au) region (Fig.\,\ref{fig:g26-spots}) within the velocity range of 3.3\kms around the peak velocity 24.2\,km\,s$^{-1}$. The maser spots are grouped into three cloudlets, with brightness temperatures ($T_{\mathrm{b}}$) ranging from $10^7$ to $4\times10^8$\,K. The spectra (\citealt{szymczak2018a}; archival data from the 32\,m telescope) at the closest dates to the EVN observation indicate that about 25\% and 70\% of the flux density in the cross-power spectrum (\citealt{bartkiewicz2009}) were missed for the two visible features at 24.2 and 24.8\,km\,s$^{-1}$, respectively, while emission at velocities of $<$22.8\kms and $>$26.0\kms were not detected at all with the synthesized beam of $6\times$12\,mas$^2$. The region of $\sim0\farcs6\times0\farcs4$ ($\sim1000\times700$\,au) was detected with VLA-C at phase 0.95 (Fig.\,\ref{fig:g26-spots}), and the cross-correlated spectrum (\citealt{hu2016}) does not indicate the missing flux effect for all eight cloudlets. The image taken at phase 0.39 with VLA-D shows the overall structure of $\sim1\farcs6\times1\farcs0$ ($\sim2800\times1800$\,au) with a velocity gradient at PA $\sim$70\degr\, for eight spots, except one isolated spot $>$1\farcs7 south of the center (Fig.\,\ref{fig:g26-spots}). The extrapolated phases of these observations, relative to the folded light curve, are marked in Fig.\,\ref{fig:g26-part-lc}b. The collected maps of maser cloudlets from the three interferometric observations are displayed in Fig.\,\ref{fig:g26-spots}. 
We conclude that the size of the source, as measured with the EVN, is a factor of 4 and 30 smaller than those determined with the VLA-C and VLA-D spatial resolutions, respectively. As the overall shape of the spectrum has remained similar since its discovery, such a significant difference in the source size can only be explained by extended emission, which is fully filtered by the beam of a few tens of milliarcseconds. Although the observations were made at different epochs and the effect of variability cannot be ruled out, they suggest the presence of compact cloudlets embedded within much larger structures of low emission intensity. It is well documented that the majority of maser regions in HMYSOs consist of compact cores surrounded by diffuse emission. Still, the extended maser areas are not always associated with compact cores (\citealt{minier2002}). They found an observational relationship that $T_{\mathrm{b}}\propto D_{\mathrm{c}}^{-2.2}$, where $D_{\mathrm{c}}$ is the diameter of core/halo diffuse emission, which is consistent with the spherical maser calculations (\citealt{emmering1994}). This relation applied to the EVN data implies a diffuse maser size of 200-700\,au in G26. 

Figure\,\ref{fig:g26-spots} shows that the cloudlets in both VLA images, where the emission does not show periodic variations in the single-dish observations, coincide within $\lesssim0\farcs1$ with the compact component detected with the EVN and the 1.3\,mm continuum source observed with ALMA. For the VLA-D map obtained 0.16$P$ before the maximum of the BS emission, the cloudlets corresponding to the 18.4 and 26.2\kms features are separated by 0\farcs96 at a position angle (PA) of 62\degr. In the VLA-C data obtained near the minimum BS emission, the distance between these regions is 0\farcs12 at PA=32\degr. We note a significant difference in the source morphology imaged with the two VLA configurations in two distinct variability phases. A clear tendency for the variable BS emission to originate in the NE region, while the variable RS emission lies in the SW region of the structure, is only seen in the VLA-D data. These distinct maser regions were not detected in the EVN observation, which suggests that the emission of the BS and RS features in the single-dish spectrum showing anticorrelated periodic variability arises in the extended regions of low maser gain.  

\begin{table*}
\centering
\caption{Properties of 1.3\,mm continuum emission.}\label{tab:table1}
\begin{tabular}{cccccc}
\hline
Component & \multicolumn{2}{c}{Position (J2000)} & Peak intensity & Flux density &  Deconvolved size \\
  & RA(s) [18$^{\mathrm{h}}39^{\mathrm{m}}$] & Dec($^{\prime\prime}$) [$-$05\degr38$^{\prime}$] & (mJy\,beam$^{-1}$) & (mJy)  & (mas$\times$mas, $\degr$)\\
\hline
C1  & 55.94295 & 44.8864 &  12.00$\pm$0.03 &  35.52$\pm$1.16 &  414.8$\times$281.9, 1.6 \\
C2  & 55.92286 & 44.7261 &  9.22$\pm$0.03 &  15.42$\pm$0.78 &  261.7$\times$141.3, 43.8 \\
\hline
\end{tabular}
\end{table*}

The position angle of $\sim$60-70\degr\, of the velocity gradient of the cloudlets is consistent with that of the elongated diffuse emission at 1.3\,mm, which extends in total about 2\farcs4 (4200\,au, Figs.\,\ref{fig:g26-spots}, \ref{fig:g26-CO-redblu}) and the CH$_3$OH thermal line at 218.4\,GHz (Fig.\,\ref{fig:g26-CH3OH-SiO}a). Within the tenuous structure, there is a compact emission elongated at PA$\approx$ $-$45\degr\, composed of two distinct cores. Their parameters derived from the Gaussian fit are listed in Table\,\ref{tab:table1}. The maser emission appears to be associated with the N-W core. The map of the SiO line (Fig.\ref{fig:g26-CH3OH-SiO}b) indicates the SE-NW velocity gradient.
The CO emission within $\sim$4500\,au of the continuum cores suggests X-shaped outflows (Fig.\,\ref{fig:g26-CO-redblu}b).

\section{Discussion}
A very peculiar feature of the G26 maser is the time lag (33\,d) of the maxima/minima of the BS emission relative to the RS one by almost half a period. Such synchronous anticorrelated and periodic variability has not been observed before. Some similarity is shown by G331.13$-$0.24 (59\,d; \citealt{goedhart2014}) and G30.400$-$0.296 (80\,d; \citealt{olech2019}), where delays between the flare peaks of the blueshifted and redshifted features were 0.12 and 0.35 phases of the period of 222 and 504\,d, respectively. Such time lags could not be explained by the light travel time between cloudlets in a typical 6.7 GHz maser region of size 300-900\,au \citep{bartkiewicz2016}. A sinusoidal-like light curve of G26 differs significantly from that of these two objects, in which the quiescent phase lasts $\sim$0.3-0.4 of the period. The intensity of BS and RS emission shows a kind of anti-cyclic beating on a timescale of $\gtrsim$ 4\,yr (Fig.\,\ref{fig:g26-spect-mm-all}b) of unknown origin, never observed in other periodic masers. We conclude that the variability pattern of the target is unusual among $\sim$30 periodic sources found so far (\citealt{goedhart2004,goedhart2014}; \citealt{maswanganye2015,maswanganye2016}; \citealt{olech2019,olech2022}; \citealt{tanabe2023}).

Synchronous and anticorrelated variability of the BS and RS components of the 6.7\,GHz spectrum was observed in Cep\,A\,HW2 \citep{sugiyama_2008, szymczak2014}. It is attributed to the changes in the dust temperature of the regions due to the different distances (300 to 900\,au, \citealt{sanna2017}) from the central star exhibiting variability. When the brightness increases, the dust temperature near the star will be too high, and the maser quenches, but it is appropriate (100-200 K) to sustain the maser at a further distance. On the other hand, decreasing would be just the opposite. In G26, the projected distance of the BS emission centroid from the putative heating source C2 (Table\,\ref{tab:table1}) is $\sim$600\,au, and that of the RS emission is $\sim$1200\,au (Fig.\,\ref{fig:g26-spots}a). The synchronous variability for such a geometry could be explained by assuming that the exciting star pulsates sinusoidally with a period of 70.1\,d, and the thermal wave induced by this propagates isotropically at a speed of 0.1c. This value of the thermal wave speed is in good agreement with that measured for an object exhibiting a 6.7\,GHz maser brightening due to an accretion outburst \citep{burns2023}. This hypothesis requires observational confirmation to determine the 3D structure of the source. Another caveat against this scenario is the lack of short-term variations in the NEOWISE W2 data greater than 0.035 magnitude. The observed changes in the maser flux are 20-40\% of the mean value; thus, for a cloudlet with $T_{\mathrm{b}} < 10^6$\,K, in the approximation of the \cite{cragg2002} model for $T_{\mathrm{k}}$ = 150\,K, the change in the maser flux density by the above value would occur when a star with an average luminosity of 3500\,$L_{\odot}$ \citep{grave2009} changes its brightness by 0.15 magnitude. 

The behavior of the maser features in G26 resembles that reported in S255 for the 22\,GHz water maser (\citealt{cesaroni1990}). It is interpreted as an effect of a radiative connection between the diametrically opposite segments of the Keplerian rotating disk seen edge-on. When the pump rate decreases on one section of the disk, the opposite section will be affected shortly after a time equal to the light crossing time. Consequently, the emission of the blueshifted features in the spectrum will vary synchronously and anticorrelate with that of the redshifted features. Still, the central features of the spectrum are relatively stable. The VLA-D observation (Fig.\,\ref{fig:g26-spots}) suggests that this scenario can be applicable for G26, but the ALMA data reported here do not provide evidence for the presence of a disk. However, a velocity gradient appears in the EW direction. The lack of evidence of a disk (typically a few hundred to 1000\,au) in the ALMA data is likely due to the low spatial resolution of the data. However, the NW-SE bipolar outflow (see Fig.\,\ref{fig:g26-CO-redblu}) with well-separated lobes points to the existence of a highly inclined circumstellar disk perpendicular to the outflow axis.

It has been postulated that the important observational parameter for resolving the variability mechanism is the shape of the light curve (e.g., \citealt{van_der_walt2011, vanderheever2019}). Fig.\,\ref{fig:g26-part-lc} implies that in the case of G26, it is similar to those observed for eclipsing stars. Among the periodic methanol sources, two objects, G338.93$-$0.06 and G358.46$-$0.39, were observed with light curves close to the absolute sine function with periods of 133 and 220\,d, respectively (\citealt{goedhart2014,maswanganye2015}). These objects have a narrow emission range ($\le$2\kms), and all features show synchronous variations by about 30-40\% of the mean flux. A simple example of a model of an eclipsing system presented by \cite{maswanganye2015} illustrates that at a distance of a few hundred astronomical units, the size of the cast shadow is on the order of tens of astronomical units.

The total luminosity of the target for the near-kinematic distance is 3500\,$L_{\odot}$ \cite{grave2009}, which suggests a mass of 15\,$M_{\odot}$. If we assume that G26 is a binary system of 8 and 7\,$M_{\odot}$ stars with an orbital period of 141\,d, then the separation between the components is 1.3\,au. In the case of eclipses, assuming a star radius of 10\,{\bf $R_{\odot}$} at typical distances of the dust disk of 300-400\,au (\citealt{sanna2017, bartkiewicz2024}), which produces pump photons $-$ the diameter of the shadow will be $\sim$21-28\,au. Therefore, a purely geometrical estimate of the shadow size suggests that the influence of eclipses, either on the flux of ionizing radiation, which affects the flux of background seed photons, or on the dust temperature, on which the pump rate depends, can be small. If the stars are bloated due to very high accretion rates, they could be even larger. In extreme cases, they could form a type of contact binary that modulates the flux by varying their surface area as they rotate.

The occurrence of rotating shadows in the disk caused by mutual occultations of binary stars has been reported in close systems of low-mass stars (e.g. \citealt{dorazi2019}). A possible alternative scenario for shadow casting is the occurrence of a misaligned/warped inner disk. These shadows are not always visible in scattered light maps. They may be caused by, among other things, central binary and disk interactions, disk interactions in triple systems (\citealt{kraus2020}), or precession of the inner parts of the disk (\citealt{debes2017}). Among these scenarios, those cases in which rotations of shadows or a moving shadow pattern have been observed may be adequate to explain the variability of the G26 maser. The 1.3\,mm continuum image of G26  (Fig.\,\ref{fig:g26-spots}) reveals at least two cores separated by 470\,au, and there may be a third. In such a system, precession of the inner disk parts is possible, but with a period two to three orders of magnitude longer than the observed variability period (\citealt{lodato2013}).

None of the simple scenarios outlined above offer a convincing explanation for the observed light curve of the 6.7\,GHz maser in G26.

\section{Conclusion}
We identify the first periodic 6.7\,GHz methanol maser source with synchronous and anticorrelated variability of the blueshifted and redshifted features. The light curves of these features have the sine shape of a period of 70.1\,d, with an amplitude of 23-36\% of the mean flux density, and they are modulated in a sine-like fashion on timescales of 8-10\,yr. A comparison of the maser maps at different angular resolutions suggests a significant flux-missing effect when observing with a 6$\times$12\,mas$^2$ beam, and the periodic anticorrelated emission originates from regions of a few thousand astronomical units, likely representing diffuse emission from halo structure.
One possible scenario we propose to explain this peculiar variability is an eclipse effect by the inner disk in a binary or multiple system, where misaligned and warped disks may occur.

The results presented here may trigger future dedicated works to advance the search for signatures of a precessing inner disk or shadows cast in a multiple system.

\begin{acknowledgements}
The 32\,m radio telescope is operated by the Institute of Astronomy, Nicolaus Copernicus University, and supported by the Polish Ministry of Science and Higher Education SpUB grant. We thank the staff and students for their assistance with the observations. 
This work was supported by the National Science Centre, Poland, through grant no. 2021/43/B/ST9/02008. JOC acknowledges support from the Italian Ministry of Foreign Affairs and International Cooperation (MAECI Grant Number ZA18GR02) and the South African Department of Science and Technology’s National Research Foundation (DST-NRF Grant Number 113121) as part of the ISARP RADIOSKY2020 Joint Research Scheme.
The research has made use of the SIMBAD database, operated at CDS (Strasbourg, France), as well as NASA's Astrophysics Data System Bibliographic Services.
This publication also makes use of data products from NEOWISE, which is a project of the Jet Propulsion Laboratory/California Institute of Technology, funded by the Planetary Science Division of the National Aeronautics and Space Administration. 

\end{acknowledgements}
%
\bibliography{librarian}{}
\bibliographystyle{aa}

\begin{appendix}
\section{Additional figures}

\begin{figure}[]
\begin{tabular}{cc}
\begin{overpic}[width=0.45\textwidth]{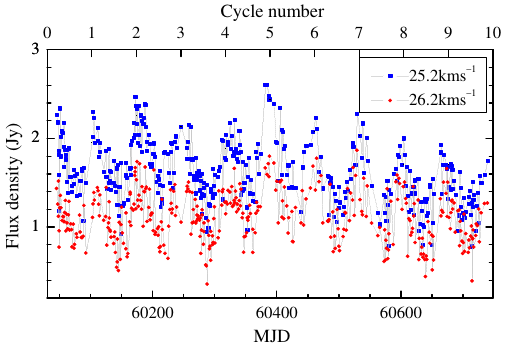}\put(11,52){(a)}\end{overpic} \\
        \begin{overpic}[width=0.45\textwidth]{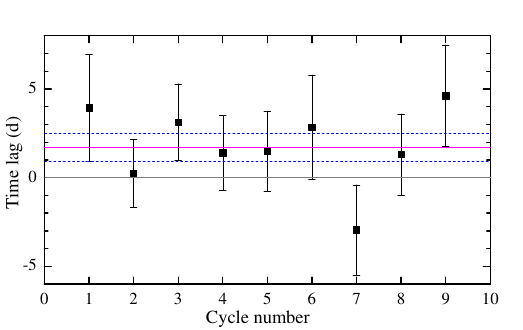}\put(11,51){(b)}\end{overpic} \\
\end{tabular}
\caption{({\bf a}) Light curves of the redshifted spectral features for MJD 60045-60739. 
({\bf b}) 
Time lag of the maximum of 26.2\kms feature relative to 25.2\kms feature for subsequent cycles. The error bars indicate the standard deviation of the data fit with the sine function. The mean lag is marked with the magenta line, and the dotted blue lines show the standard error.
   \label{fig:g26-timelag}}
\end{figure}

\FloatBarrier
\begin{figure}
\includegraphics[width=0.45\textwidth]{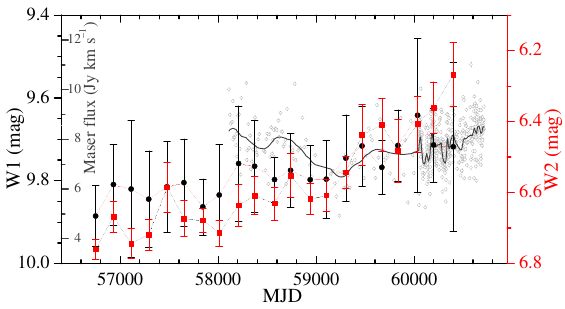}
\caption{6.7\,GHz maser velocity-integrated flux density (grey labels and circles) of G26 superimposed with the 3.4\,$\mu$m (W1) and 4.6\,$\mu$m (W2) NEOWISE light curves. The average magnitudes are shown for 21 NEOWISE epochs, and the bars denote the standard deviation. The dark grey line shows the average maser light curves.
   \label{fig:g26-neowise}}
\end{figure}

\begin{figure*}
   \begin{tabular}{cc}
\begin{overpic}[width=0.5\textwidth]{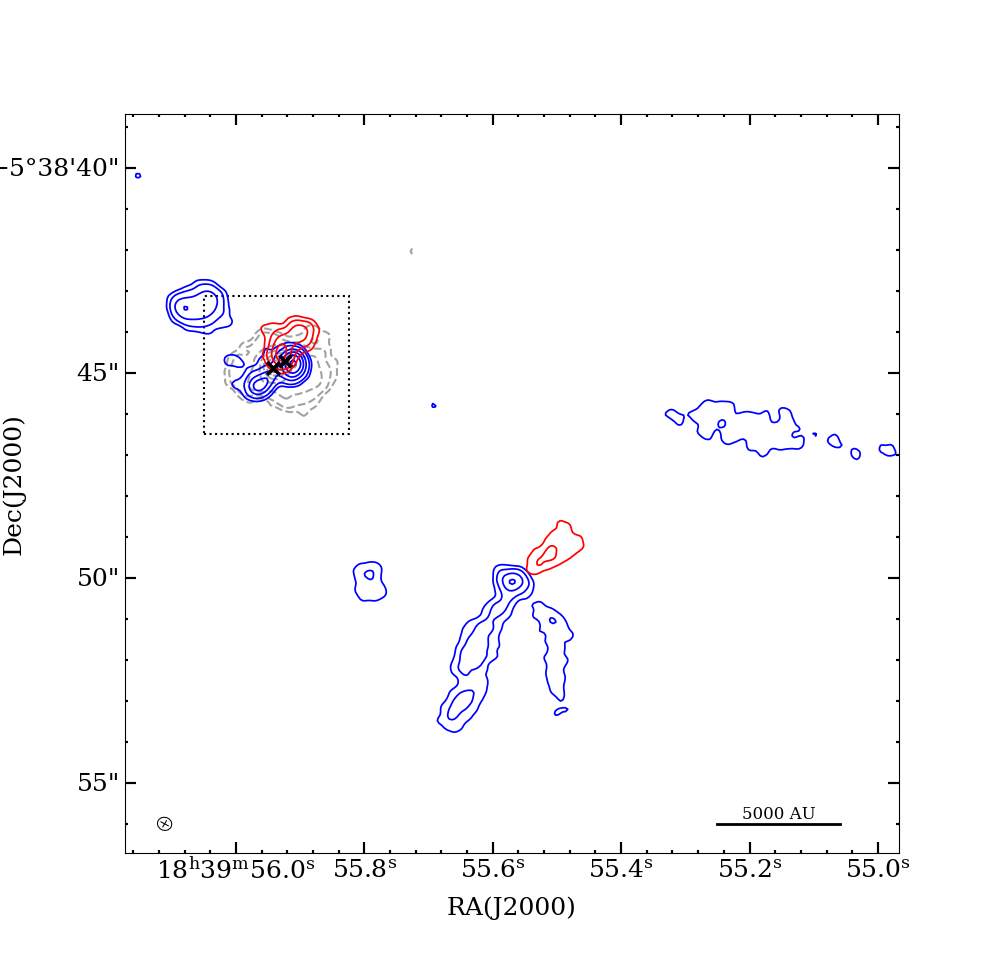}%
\put(14,80){(a)}%
\end{overpic}%
~%
\begin{overpic}[width=0.5\textwidth]{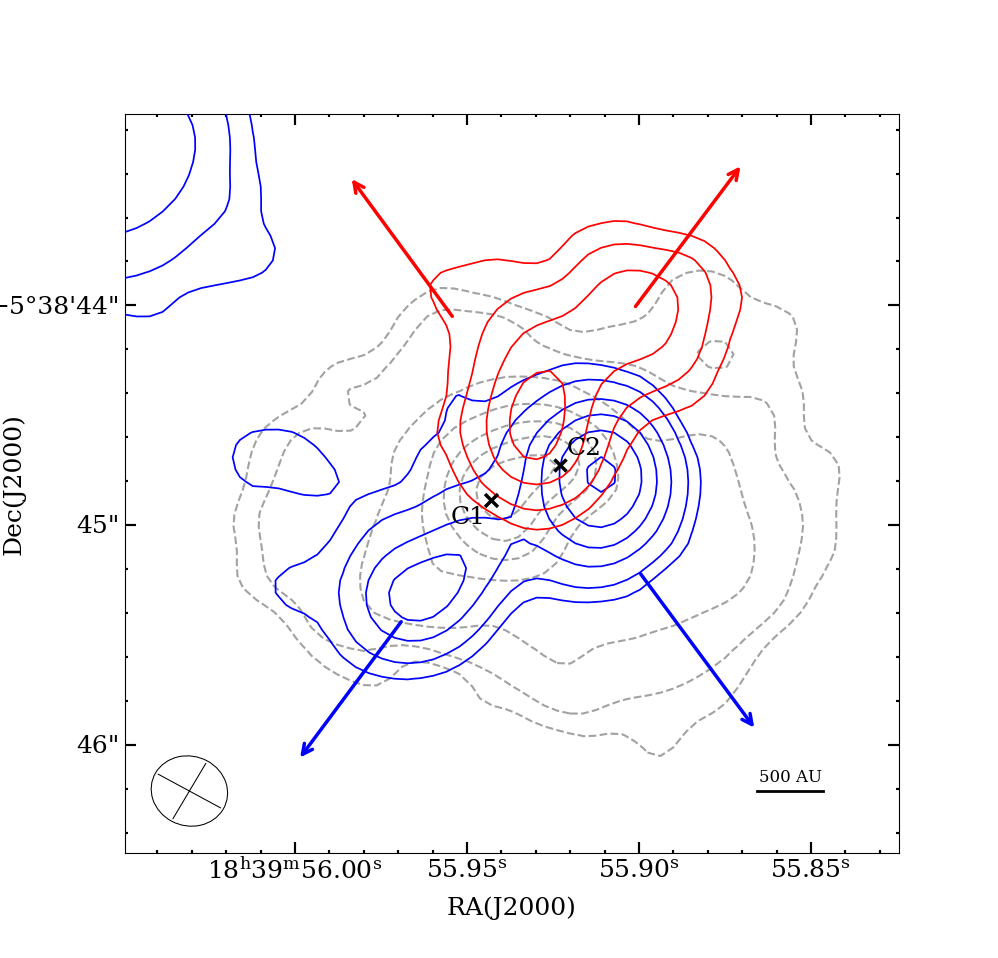}%
\put(14,80){(b)}%
\end{overpic}%
\end{tabular}
     \caption{({\bf{a}}) Overview of superposition of the 1.3\,mm continuum (grey dashed contours) and CO (2-1) blueshifted and redshifted (blue and red contours, respectively) emission toward G26.  Contours are 3, 5, 10, 20, 30, 40, 50 $\times 1\sigma_{\mathrm{rms}}$ of 0.3\,mJy\,beam$^{-1}$ for the continuum and 5, 10, 20, 30, 40, 50, 55 $\times 1\sigma_{\mathrm{rms}}$ of 3.0\,mJy\,beam$^{-1}$ for CO. The redshifted and blueshifted emissions are integrated between {37} and {46}\kms and between {7} and {17}\kms with respect to the systemic velocity of 24.4\,km\,s$^{-1}$, respectively. The dashed square marks the region enlarged in ({\bf{b}}). The crosses represent the cores labelled with C1 and C2, whose parameters are listed in Table\,\ref{tab:table1}. The blue and red arrows correspond to tentative directions of bipolar blueshifted and redshifted molecular outflows. The synthesized beam is shown in the bottom left for both images.}
    \label{fig:g26-CO-redblu}
\end{figure*}

\begin{figure*}
   \begin{tabular}{cc}
\begin{overpic}[width=0.5\textwidth]{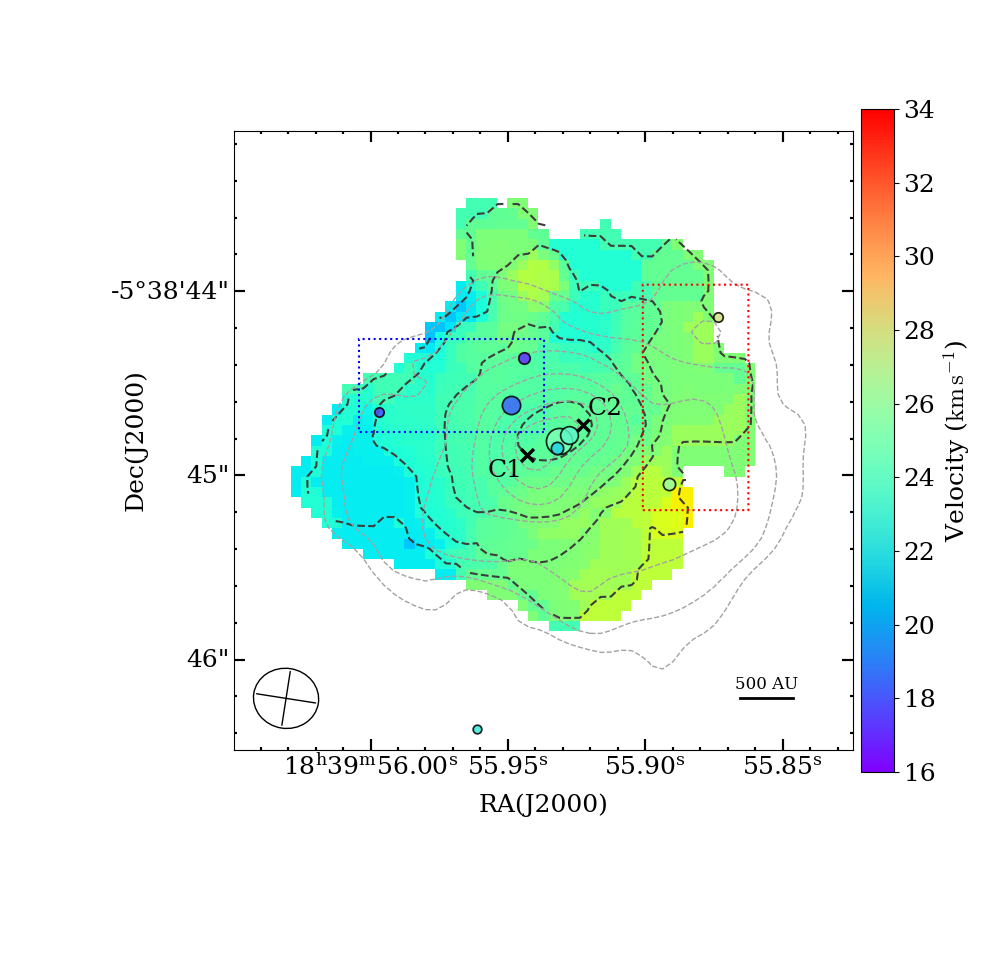}%
\put(18,71){(a)}%
\end{overpic}%
~%
\begin{overpic}[width=0.5\textwidth]{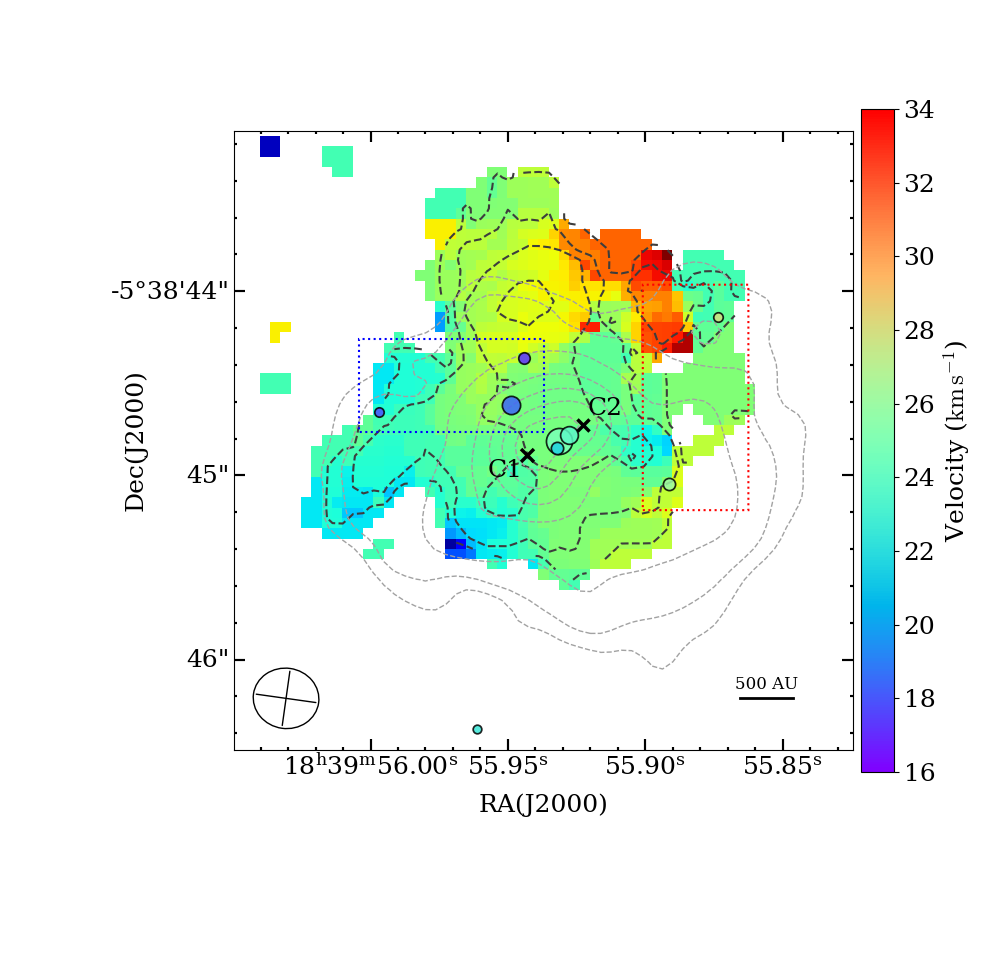}%
\put(18,71){(b)}%
\end{overpic}%
\end{tabular}
     \caption{Moment 0 (dashed black contour) and 1 (color scale) maps of the CH$_3$OH 218.440\,GHz ({\bf{a}}) and SiO 217.205\,GHz ({\bf{b}}). The 1.3\, mm continuum emission is shown with the grey dashed contours of the same levels as in Fig.\ref{fig:g26-CO-redblu}. The contours for CH$_3$OH and SiO are 5, 15, 30, 90 times 1$\sigma_{\mathrm{rms}}$ of 11.70\,mJy\,beam$^{-1}$\kms and 3, 7, 15, 30 times 1$\sigma_{\mathrm{rms}}$ of 15.49\,mJy\,beam$^{-1}$\kms, respectively. The maser cloudlets taken from \cite{nguyen2022} are represented with circles whose sizes are proportional to the logarithm of brightness and colors scale to the velocity shown by the wedges. The cloudlets are grouped by colored, dashed boxes corresponding to the BS and RS emission, showing periodic anticorrelated variability, as defined in Sect.\,\ref{sec:results}. 
     The synthesized beam is shown in the bottom left for both images.}
    \label{fig:g26-CH3OH-SiO}
\end{figure*}

\end{appendix}

\end{document}